\documentstyle[12pt,epsfig]{article}
\begin{document}

\title{
Measurement of the R$_{LT}$ response function for $\pi^0$ electroproduction
at $Q^2$ = 0.070 (GeV/c)$^2$ in the $N\rightarrow\Delta$ transition}

\date{ \small (\today)}

\maketitle

\author{
N.F.~Sparveris$^1$, R.~Alarcon$^2$, D.~Barkhuff$^4$,  A.~Bernstein$^4$, W.~Bertozzi$^4$, 
J.~Calarco$^6$, F.~Casagrande$^4$, J.~Chen$^4$, J.~Comfort$^2$, 
M.O.~Distler$^4$, G.~Dodson$^4$, A.~Dooley$^7$, K.~Dow$^4$, 
M.~Farkondeh$^4$, 
S.~Gilad$^4$, R.~Hicks$^5$, M.~Holtrop$^6$, A.~Hotta$^5$, X.~Jiang$^5$, N.~Kaloskamis$^{1,4}$, 
A.~Karabarbounis$^1$, S.~Kowalski$^4$, C.~Kunz$^4$, D.~Margaziotis$^3$, C.~Mertz$^1$, R.~Milner$^4$, 
R.~Miskimen$^5$, I.~Nakagawa$^4$, 
C.N.~Papanicolas$^1$, 
M.M.~Pavan$^4$, G.~Peterson$^5$, A.~Ramirez$^2$, 
D.~Rowntree$^4$, A.J.~Sarty$^7$, J.~Shaw$^5$, E.~Six$^2$, S.~Soong$^4$, S.~Stiliaris$^1$, 
D.~Tieger$^4$, C.~Tschalaer$^4$, 
\centerline{W.~Turchinetz$^4$, C.~Vellidis$^1$, G.A.~Warren$^4$, A.~Young$^2$, J.~Zhao$^4$,} \\
\centerline{Z.-L.~Zhou$^4$ and T.~Zwart$^4$}
\\
\\
\centerline{$^1$Institute of Accelerating Systems and Applications} \\ 
\centerline{and University of Athens, Athens, Greece}   \\
\centerline{$^2$Department of Physics and Astronomy, Arizona State University,}  \\ 
\centerline{$^3$Department of Physics, California State University,}   \\
\centerline{$^4$Department of Physics and Bates Linear Accelerator Center,}  
\centerline{Massachusetts Institute of Technology,}   \\
\centerline{$^5$Department of Physics, University of Massachusetts,}   \\
\centerline{$^6$Department of Physics, University of New Hampshire,}  \\
\centerline{$^7$Department of Physics, Florida State University}
}

\newpage

\begin{abstract}
Quadrupole amplitudes in the $\gamma^{*}N\rightarrow\Delta$ 
transition are associated with the issue of nucleon deformation. 
A search for these small amplitudes has been the focus of a series of 
measurements undertaken at Bates/MIT by the OOPS collaboration. We report on results 
from H$(e,e^\prime p)\pi^0$ data obtained at $Q^2= 0.070$ (GeV/c)$^2$ and invariant 
mass of $W$=1155 MeV using the out-of-plane detection technique with the OOPS 
spectrometers. The $\sigma_{LT}$ and $\sigma_{T}$+$\epsilon\cdot$$\sigma_{L}$ response 
functions were isolated. These results, along with those of previous measurements at 
$W$=1172 MeV and $Q^2= 0.127$ (GeV/c)$^2$, aim in elucidating the interplay between 
resonant and non resonant amplitudes.

\end{abstract}

\section{Introduction}

The signature of the conjectured deformation of the nucleon \cite{gla79}
is mostly saught through the isolation of resonant quadrupole amplitudes 
in the $\gamma^* N\rightarrow \Delta$ transition. Such quadrupole
contributions provide a sensitive probe of the
internal nucleon structure and the underlying quark dynamics.
Quadrupole amplitudes and the origin of deformation is attributed to
different effects depending on the theoretical approach adopted.
In a constituent-quark picture of the nucleon, a quadrupole resonant 
amplitude would point to a d-state admixture in the 3-quark wave function 
of the nucleon.  Such a d-state component is expected as a consequence of 
a spin-spin tensor or color-hyperfine interaction among quarks.
In dynamical models of the $\pi N$ system, the effect of the pionic cloud 
will also allow the appearance of quadrupole amplitudes.

A number of experimental programs \cite{pho1,pho2,bart,joo,frol,pos01,goth} 
have been active in photo- and electro-pion
production in the $\Delta$ region at all the itermediate energy electromagnetic
facilities. Results emerging from these programs strongly support the notion of 
a deformed nucleon although the magnitude and the origin of this effect is still 
under exploration. The principal difficulty derives from the "contamination"
of the quadrupole amplitudes from coherent processes, such as Born terms or tails
of higher resonances. The isolation of the contributions of the non resonant
terms has emerged as a key task in the experimental program exploring the issue 
of nucleon deformation. Recent reviews on this issue can be found in 
\cite{nstar,rev2,rev3}.

We present here the results pertaining to the measurement of the 
$\sigma_{LT}$ and $\sigma_{o}=\sigma_{T}$+$\epsilon\cdot$$\sigma_{L}$ 
response functions in an H$(e,e^\prime p)\pi^0$ reaction at $Q^2= 0.070$ 
(GeV/c)$^2$ and at $W$=1155 MeV, on the rising shoulder of the $\Delta$ resonance.
The motivation for the experiment was twofold:
a) to understand the interplay between resonant and non resonant amplitudes
which is best explored by following the $W$ dependence of the various responses
and b) to commence a series of measurements at a lower $Q^2$ than 
$Q^2= 0.127$~(GeV/c)$^2$, a point where the database
is by now quite rich as a result of measurements at Bates, Mainz and Bonn 
\cite{pos01,merve,vellthes,kal97}. The reason of focusing on
low momentum transfer region is driven by the need to understand the pion 
cloud effects which are expected to dominate the E2 and C2 transition matrix 
elements in the low $Q^2$ (large distance) scale.

Spin-parity selection rules in the $N(J^{\pi}=1/2^+)\rightarrow \Delta(J^{\pi}=3/2^+)$ 
transition, allow only magnetic dipole $(M1)$ and electric quadrupole $(E2)$ 
or Coulomb quadrupole $(C2)$ multipoles to contribute.
The resonant photon absorption multipoles $M1, E2$, and $C2$ 
correspond to the pion production multipoles $M^{3/2}_{1+}$, $E^{3/2}_{1+}$, 
and $S^{3/2}_{1+}$, respectively, following the notation
$M^{I}_{l\pm }$, $E^{I}_{l\pm }$, and $S^{I}_{l\pm }$, 
where $I$ and $J= l\pm\frac{1}{2}$ correspond to their isospin
and orbital angular momentum respectively.  
The quadrupole amplitudes are typically referred to in terms of their ratio to the
dominant magnetic dipole amplitude $M^{3/2}_{1+}$. The Coulomb quadrupole to Magnetic
Dipole Ratio is defined as $CMR = R_{SM} = Re(S^{3/2}_{1+}/M^{3/2}_{1+})$ and the Electric
quadrupole to Magnetic Dipole Ratio as $EMR = R_{EM} = Re(E^{3/2}_{1+}/M^{3/2}_{1+})$.
In the spherical quark model of the nucleon, the 
$N \rightarrow \Delta$ excitation is a pure $M1$ transition. 
Models of the nucleon which are in reasonable agreement with the known
experimental facts \cite{sato,mai00,kama,dmt00} predict values of $R_{SM}$ in the range 
of -1\% to -7\%, at momentum transfer square $Q^2 \approx  0.1$~(GeV/c)$^2$.

The fact that the amplitudes of interest contribute only to a very small fraction of
the reaction cross section leads to the conclusion that the most sensitive responses,
the ones that carry the signal of our interest, will be interference responses
in which the weak quadrupole amplitudes will manifest themselves through interference
with the dominant dipole amplitude. The interference of the $C2$ amplitude with
the $M1$ will obviously lead to Longitudinal~-~Transverse (LT) type responses.
The determination of the $\sigma_{LT}$ response was the primary target of this experiment.

\section{Experimental method}

The cross section of the H$(e,e^\prime p)\pi^0$ reaction is sensitive to
four independent response functions:

\begin{eqnarray}
 \frac{d^5\sigma}{d\omega d\Omega_e d\Omega^{cm}_{pq}} & = & \Gamma (\sigma_{T} + \epsilon{\cdot}\sigma_L
  - v_{LT}{\cdot}\sigma_{LT}{\cdot}\cos{\phi_{pq}} \nonumber \\
 & &  +\epsilon{\cdot}\sigma_{TT}{\cdot}\cos{2\phi_{pq}} )
\label{equ:cros}
\end{eqnarray}
where the kinematic factor 

\begin{eqnarray}
 v_{LT}=\sqrt{2\epsilon(1+\epsilon)} \nonumber 
\label{equ:fact}
\end{eqnarray}
and $\epsilon$ is the transverse polarization of the virtual photon, $\Gamma$ the virtual
photon flux and $\phi_{pq}$ is the proton azimuthal angle with respect to the momentum
transfer direction.

In the experiment reported here we have measured the $\sigma_{LT}$ response function, which
contains the interference term $Re(S^{*}_{1+}M_{1+})$ in leading order \cite{multi},
and the $\sigma_{T}$+$\epsilon\cdot$$\sigma_{L}$ which is dominated by the $\sigma_{T}$ response
and the $M_{1+}$ multipole . The measurement was performed using the technique of the out-of-plane 
detection with the OOPS spectrometers. By placing the two identical OOPS modules \cite{oopsdolf,oopsmand} 
symmetrically at azimuthal angles $\phi_{pq}= 45^o$ and $135^o$ with respect to the momentum 
transfer direction - in the so called "half-{\large $\times$} configuration" - we have 
the advantage of eliminating out in leading order the $\sigma_{TT}$ response term from the 
cross section because of its $\cos{2\phi_{pq}}$ dependence. Thus, combining the measurements from 
the two OOPS spectrometers we are able to separate the  $\sigma_{LT}$ and 
$\sigma_{T}$+$\epsilon\cdot$$\sigma_{L}$ responses
(eq.~\ref{equ:rlt} and \ref{equ:rt}). The A$_{LT}$ asymmetry is also measured 
(eq.~\ref{equ:alt}) which is proportional to the $\sigma_{LT}$ response 
and inversely proportional to  $\sigma_{T}$+$\epsilon\cdot$$\sigma_{L}$

\begin{equation}
\sigma_{LT}    ~ = ~ \frac{1}{\sqrt{2}{\cdot}v_{LT}}
\left[ \frac{d^2\sigma}{d\Omega^{cm}_p}(\phi_{pq}=\frac{\pi}{4}) - 
 \frac{d^2\sigma}{d\Omega^{cm}_p}(\phi_{pq}=\frac{3\pi}{4}) \right]
\label{equ:rlt}
\end{equation}

\begin{equation}
\sigma_{T}+\epsilon{\cdot}\sigma_{L}    ~ = ~ \frac{1}{2}
\left[ \frac{d^2\sigma}{d\Omega^{cm}_p}(\phi_{pq}=\frac{\pi}{4}) + 
 \frac{d^2\sigma}{d\Omega^{cm}_p}(\phi_{pq}=\frac{3\pi}{4}) \right]
\label{equ:rt}
\end{equation}

\begin{eqnarray}
A_{LT}=&&\frac{d\sigma(\phi_{pq} = \pi/4) - d\sigma(\phi_{pq} = 3\pi/4)}
                  {d\sigma(\phi_{pq} = \pi/4) + d\sigma(\phi_{pq} = 3\pi/4)}\nonumber\\
&&=  \frac{v_{LT}{\cdot}\sigma_{LT}}{\sqrt2 (\epsilon{\cdot}\sigma_{L} + \sigma_{T})}
\label{equ:alt}
\end{eqnarray}

The fact that the measurements were performed simultanously with two 
identical proton spectrometers enabled us to minimize the systematic errors. 
Minimization of the systematic errors is a key issue in this experiment since 
the quadrupole amplitude of interest contributes only as a very small part of 
the reaction cross section.

\section{Experiment and results}

The experiment was performed in the South Hall of M.I.T.-Bates Laboratory. A 0.85\% duty
factor, 820 MeV unpolarized pulsed electron beam was employed on a cryogenic 
liquid-hydrogen target. The beam average current was 5 $\mu A$. Protons were detected 
with two OOPS spectrometers \cite{oopsdolf,oopsmand,oopsnim}. They were  symmetrically positioned at 
$\phi_{pq}= 45^o$ and $135^o$ with respect to the momentum  transfer direction for a 
fixed  $\theta_{pq}^{*}$=$55^o$ and were set at a central momentum of 428 MeV/c.
The uncertainty in the determination of the central momentum was 0.1\% for the proton
arm and 0.15\% for the electron arm. Electrons were detected with the OHIPS spectrometer 
\cite{xia98} which was located at an angle of 22.9$^{o}$ and was set at a central momentum of 541 MeV/c.
The uncertainty in the determination of the beam energy was 0.1\%. The spectrometers
were aligned with a precision better than 1 mm and 1 mrad, while the uncertainty in the
determination of the total beam charge was 0.1\%.
The central invariant mass and the squared four-momentum transfer were $W$ = 1155 MeV and 
$Q^2$ =  0.070 GeV$^2$/c$^2$ respectively.
A third OOPS was set to detect elastically scattered electrons and was used as a luminosity monitor 
throughout the experiment. It was placed in-plane at an angle of 75.8$^{o}$ and was set at a central
momentum value of 494 MeV/c.

The OHIPS spectrometer employed two Vertical Drift Chambers for the track
reconstruction. Two layers of 14 Pb-Glass detectors and a Cherenkov detector were 
responsible for identification of electrons from the $\pi^{-}$ backgound.
The timing information for OHIPS derived from 3 scintillator detectors. 
The OOPS spectrometers used three Horizontal Drift Chambers for the track 
reconstruction followed by three scintillator detectors for timing and for the
separation of the protons from the strong $\pi^{+}$ background coming from the
$\gamma^* p \rightarrow \pi^{+} n$ processes on hydrogen in the target. 

The data taking period was preceded by a comissioning period. 
Elastic scattering data for calibration purposes were taken using liquid-hydrogen 
and carbon targets and a 600 MeV beam. Measurements were conducted with and without 
sieve slits in all spectrometers. The sieve slit runs were used to determine the 
optical matrix elements for all spectrometers \cite{kun00}, while the runs without 
sieve slits were used for the elastic cross sections and normalization studies.

The "on line" coincidence time-of-flight peak had a FWHM of 6 ns.  
After the time-of-flight corrections were applied to account for differences in
the particle path length, particle velocities, different light-times in the 
scintillators and time walk effects in the scintillators, the FWHM was reduced to 2 ns. 
The missing mass spectrum with the peak corresponding to the reconstruction of the 
$\pi^{0}$ mass was characterized by a width of 8 MeV (FWHM) and was succesfully
simulated by the Monte Carlo simulation. 
A cut of a 5.5 ns time window on the corrected time-of-flight and of $\pm$10 MeV 
around the missing mass peak was used to select good events throughout the analysis.

The Monte Carlo program AEEXB \cite{vel98} was used to model the experimental setup. 
A detailed simulation of the spectrometers involved was necessary in order to
determine the coincidence phase space volume. The precise knowledge of this
volume was essential for the determination of absolute cross sections.

The conventional set of three independent kinematic variables that is used to
describe the cross section is the center-of-mass opening angle, the four
momentum transfer squared and the invariant mass \{$\theta_{pq}^*$,$Q^2$,$W$\}.
Extraction of A$_{LT}$ and $\sigma_{LT}$ requires that the phase space
of the detected protons is identical in the two OOPS spectrometers. 
However, due to the extended acceptances and the different convolution with the 
electron acceptance, the accessible range in these three variables differ for the
two proton arms. For this reason, the cross sections were measured individually for 
the two spectrometers with their respective coincident phase space volumes
\{$\theta_{pq}^*$,$Q^2$,$W$\} matched.
To facilitate comparison with theoretical predictions, we corrected our
measured cross sections for finite acceptance effects using theoretical
models by comparing the model cross section for point kinematics to the same
model averaged over the full acceptance.

The cross sections for the {\it forward} (with respect to the beam) OOPS $(\phi_{pq}=\frac{\pi}{4})$ 
and the {\it backward} OOPS $(\phi_{pq}=\frac{3\pi}{4})$, along with the A$_{LT}$ space asymmetry 
and the responses $\sigma_{LT}$ and $\sigma_{T}$+$\epsilon{\cdot}$$\sigma_{L}$ are summarized in 
Table~\ref{tab:resu}. The results are compared with the recent theoretical models 
MAID 2000 \cite{mai00,kama} and the Dynamical Models of DMT (Dubna - Mainz - Taipei) \cite{dmt00} 
and Sato-Lee \cite{sato}. Results from these models have been widely used in comparisons with 
recent experimental results. We will therefore forego a summary of their physical content which 
is presented in the original papers and other recent experimental investigations.

\begin{table}[h]\centering
\begin{tabular}{|l|cl|} \hline
       $W$                                                		  &  1155             & MeV         \\
       $Q^2$                                            		  &  0.070            &(GeV/c)$^2$  \\
       $\theta^{*}_{pq}$                                		  &  $55^{\circ}$     &             \\\hline
{\large  $\frac{d\sigma}{d\Omega}$ } $(\phi_{pq}=\frac{\pi}{4})$   	  & 9.72$\pm$0.46     & $\mu$b/sr    \\
{\large  $\frac{d\sigma}{d\Omega}$ } $(\phi_{pq}=\frac{3\pi}{4})$	  & 11.05$\pm$0.49    & $\mu$b/sr    \\
       A$_{LT}$               						  &  -6.4$\pm$2.4     & \%           \\
       $\sigma_{LT}$          				       		  &   0.53$\pm$0.19   & $\mu$b/sr    \\
       $\sigma_{o}=\sigma_{T}$+$\epsilon{\cdot}$$\sigma_{L}$	 	  &  10.39$\pm$0.31   & $\mu$b/sr \\ \hline
\end{tabular} 
\caption{Table of results.}
\label{tab:resu}
\end{table}

\begin{figure}[h]
\centerline{\psfig{figure=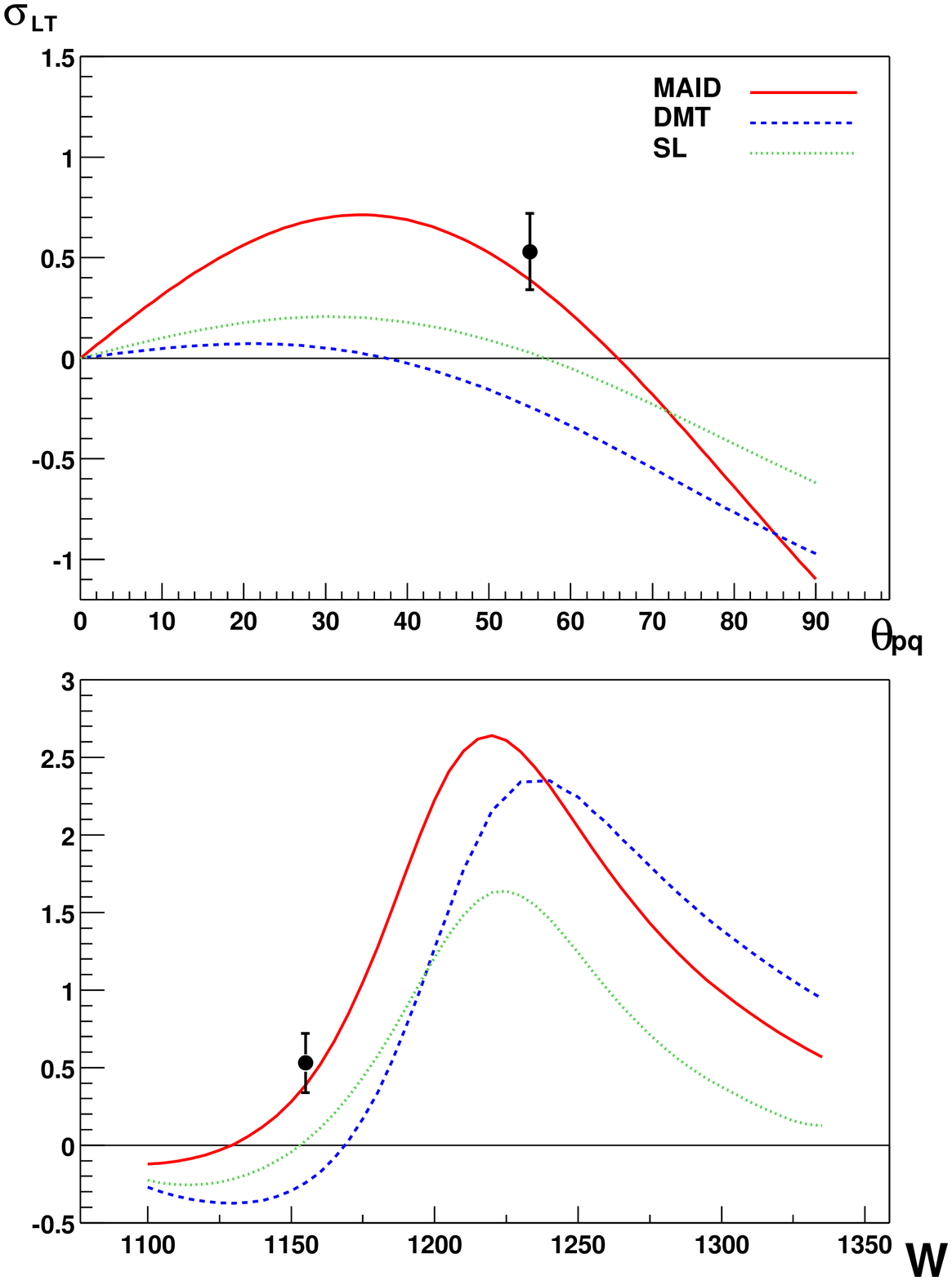,width=10.0cm,height=16.70cm,angle=000}}
\label{mrlt}
\smallskip
\caption{The $\sigma_{LT}$ response function measured at this experiment
is plotted as a function of $\theta_{pq}^{*}$ (top) and as a function of
the invariant mass $W$ (bottom) along with the predictions 
of the MAID, DMT and Sato-Lee model calculations}
\label{fig:rlt}
\end{figure}

\begin{figure}[h]
\centerline{\psfig{figure=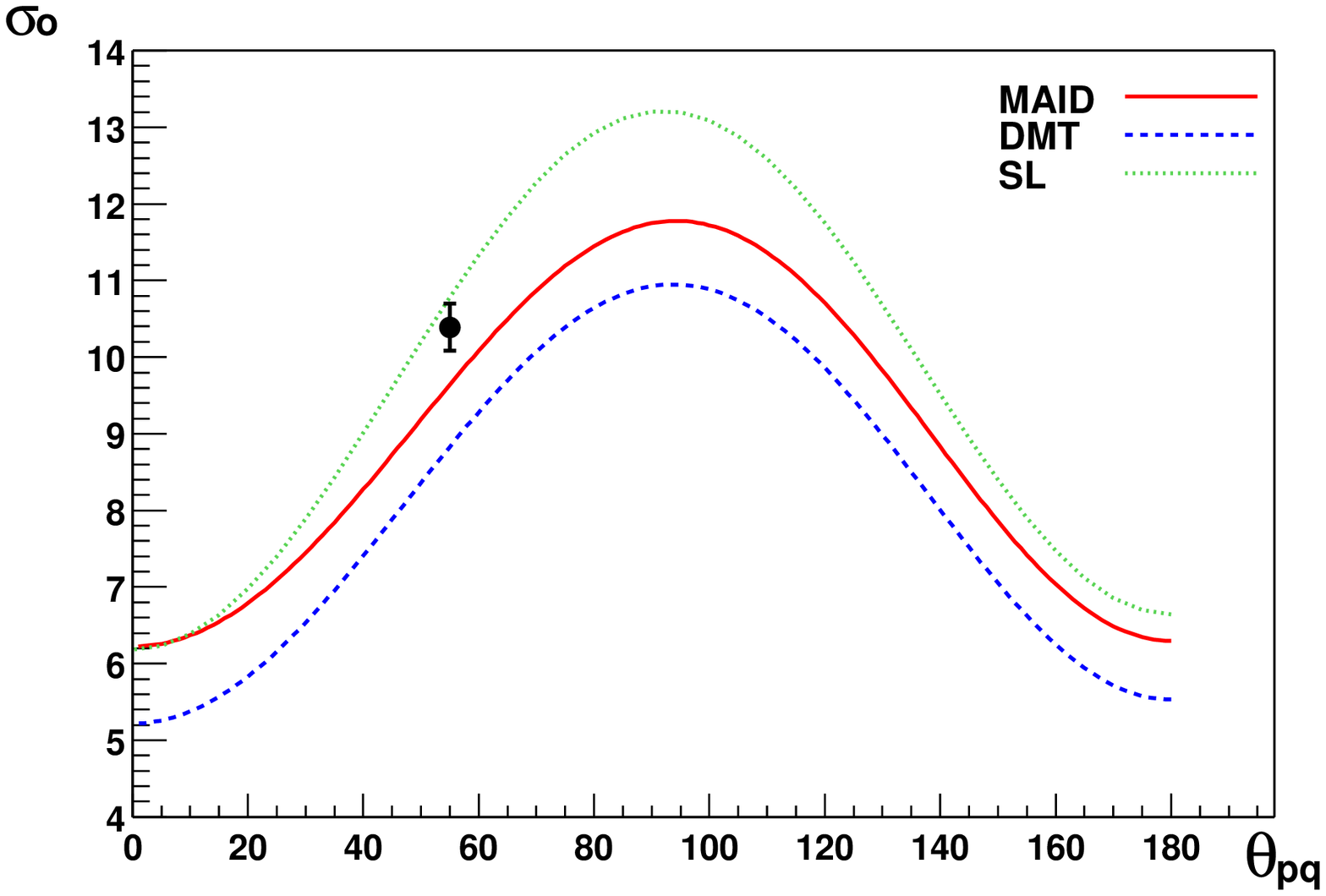,width=10.0cm,height=8.35cm,angle=000}} 
\label{rt}
\smallskip
\caption{The $\sigma_{o}=\sigma_{T}$+$\epsilon{\cdot}$$\sigma_{L}$ responses sum measured
at this experiment is plotted as a function of $\theta_{pq}^{*}$ along with the model 
calculations.}
\label{fig:rt}
\end{figure}

In Figures~\ref{fig:rlt} and \ref{fig:rt} we present the experimental
results for $\sigma_{LT}$ and $\sigma_{T}$+$\epsilon{\cdot}$$\sigma_{L}$ along
with the above mentioned model calculations \cite{sato,mai00,dmt00}. The MAID 2000 model offers 
consistently the best description of the data obtained so far \cite{merve,kun00,ware} 
with a slight tendency to somewhat underpredict the strength of the measured 
responses. Surprisingly, the DMT calculation which had considerable success in describing data 
on resonance (at higher $Q^2$ values), is incompatible with the experimental results 
presented here. The Sato-Lee model calculation, which as the DMT offers an economic 
phenomenological description anchored in a consistent microscopic framework, similarly 
underpredicts the $\sigma_{LT}$ response with results straggling the difference between 
the MAID and DMT models. The idadequacy of the dynamical models but also their differences
suggest that they may be capable of describing the data in a more satisfactory fashion
with a re-adjustment of their phenomenological input. It is evident from both figures
that data of similar or higher precison at several $\theta_{pq}^{*}$ and $W$s are
needed to further enhance our understanding of the dificiencies of these models.

Our earlier measurements \cite{merve,kun00} below the $\Delta$ 
resonance - at $W$=1170 MeV, $Q^2$= 0.127 (GeV/c)$^2$ and at $\theta_{pq}^{*}$=$61^o$ - 
exhibit a similar trend when compared with the predictions from these models.
MAID 2000, which provides an excellent account of the measured responses on resonance 
\cite{merve,vellthes,kun00}, offers a good description of the measured responses below 
resonance \cite{merve,vellthes,kun00} with a tendency to slightly underpredict them. 
This may be due to multipoles that are not well determined 
in the model and which play a relatively more important role away from the peak of 
the resonance. The dynamical models, DMT and Sato-Lee, clearly exhibit 
deficiencies off resonance. These deficiencies taken together with the behaviour of the models 
on top of the resonance \cite{merve,vellthes,kun00} indicate that the dynamic of models 
need further refinement in order to account for the delicate interplay between non
resonant and resonant amplitudes, which is manifested most sensitively at the wings 
of the resonance. A better understanding of the interfering amplitudes and their isolation 
can be facilitated through an extensive and detailed mapping of the responses
primarily in terms of $W$ and $\theta_{pq}^{*}$, as it is evident from figures~\ref{fig:rlt} 
and \ref{fig:rt}, but also in terms of $Q^2$.

Recent measurements \cite{sparve} at $Q^2= 0.127$ $(GeV/c)^2$, on and above
resonance utilizing the OOPS spectrometers are currently being analyzed 
and are expected to provide a more complete picture of the behaviour of 
the responses of the $\pi^0$ electroproduction in the $N\rightarrow\Delta$ 
transition; They are expected to elucidate further issues related to hadron deformation.

We are indebted and would like to thank Dr S.S.~Kamalov, T.-S.H.~Lee, L.~Tiator and T.~Sato 
for providing us with valuable suggestions on the overall program and these results in particular.

%%\newpage

%-----------------------------------------------------------------------------
%-----------------------------------------------------------------------------

%\newpage

%-----------------------------------------------------------------------------
%-----------------------------------------------------------------------------


\begin{thebibliography}{}
\bibitem{gla79} S.L.~Glashow, {\it Physica} {\bf 96A}, 27 (1979).

\bibitem{pho1} R.~Beck {\it et al}. {\it Phys. Rev.} C {\bf 61}, 35204 (2000).

\bibitem{pho2} G.~Blanpied {\it et al}. {\it Phys. Rev. Lett.} {\bf 79}, 4337 (1997).

\bibitem{bart} P.~Bartsch {\it et al}. {\it Phys. Rev. Lett.} {\bf 88}, 142001 (2002).

\bibitem{joo} K.~Joo {\it et al}. {\it Phys. Rev. Lett.} {\bf 88}, 122001 (2002). 

\bibitem{frol} V.V.~Frolov {\it et al}. {\it Phys. Rev. Lett.} {\bf 82}, 45 (1999). 

\bibitem{pos01} T. Pospischil {\it et al}. {\it Phys. Rev. Lett. {\bf 86} (2001), 2959}.

\bibitem{goth} R.~Gothe contribution to \cite{nstar}.

\bibitem{nstar} See e.g. NStar 2001, Proceedings of the Workshop on the Physics of
Excited Nucleons, D. Drechsel and L. Tiator editors, World Scientific (2001).

\bibitem{rev2} C.N. Papanicolas, International Conference on Quark Nuclear Physics, Julich, Germany,
June 9-14 (2002) , to be published.  

\bibitem{rev3} A.M. Bernstein, Electron-Nucleus Scattering VII Elba, Italy June 23-28 (2002),
to be published. 

\bibitem{merve} C.~Mertz {\it et al}. {\it Phys. Rev. Lett.} {\bf 86}, 2963 (2001).

\bibitem{vellthes} C.~Vellidis, Ph.D. thesis,  University of Athens, Greece (2001),
ISBN: 960-8313-05-8.

\bibitem{kal97} F. Kalleicher {\it et al.}, {\it Z. Phys.}, 201-204 (1997).

\bibitem{sato} T.~Sato and T.-S.H.~Lee, {\it Phys. Rev.} C {\bf 63}, 055201 (2001) ;
T.~Sato and T.-S.H.~Lee, {\it Phys. Rev.} C {\bf 54}, 2660 (1996).

\bibitem{mai00} D.~Drechsel {\it et al.}, {\it Nucl. Phys.} {\bf A645}, 145 (1999)
and http://www.kph.uni-mainz.de/MAID/ maid2000/maid2000.html.

\bibitem{kama} S.S. Kamalov {\it et al.}, {\it Phys. Lett. {\bf B 522} (2001), 27}.

\bibitem{dmt00} S.S. Kamalov and S.N. Yang, {\it Phys. Rev. Lett.} {\bf 83}, 4494 (1999)
and http://www.kph.uni-mainz.de/ MAID/dmt/dmt2001.html.

\bibitem{multi} D.~Drechsel and L.~Tiator, {\it J. Phys.} {\bf G18}, 449 (1992)

\bibitem{oopsdolf} S.M. Dolfini {\it et al.}, {\it Nuclear Inst. and Meth.} {\bf A 344}, 571 (1994).

\bibitem{oopsmand} J. Mandeville {\it et al.}, {\it Nuclear Inst. and Meth.} {\bf A 344}, 583 (1994).

\bibitem{oopsnim} Z. Zhou {\it et al.}, {\it Nuclear Inst. and Meth.} {\bf A 487}, 365-380 (2002).

\bibitem{xia98} X.~Jiang, Ph.D. thesis, University of Massachusetts (1998).

\bibitem{kun00} C.~Kunz, Ph.D. thesis, Massachusetts Institute of Technology (2000),
to be published.

\bibitem{vel98} C.~Vellidis, {\it AEEXB, A program for Monte Carlo simulations of coincidence
electron scattering experiments}, MIT/Bates internal report (1998).

\bibitem{ware} G.~Warren  {\it et al.}, {\it Phys. Rev.} C {\bf 58}, 3722 (1998).

\bibitem{sparve} N.~Sparveris, Ph.D. thesis, University of Athens, Athens, Greece (2002) ; 
S.~Georgakopoulos, Ph.D. thesis, University of Athens, Athens, Greece, in preparation.

\end{thebibliography}
\end{document}